\documentclass[twocolumn,aps,prl,floatfix,superscriptaddress]{revtex4-1}

\usepackage{hyperref}
\usepackage[utf8]{inputenc}
\usepackage{newunicodechar}
\newunicodechar{ }{\,}

\usepackage{graphicx}
\usepackage{dcolumn}
\usepackage{amsmath}
\usepackage{supertabular}
\usepackage{multirow}

\usepackage{lineno}

\usepackage{color}
\usepackage{cancel}
\usepackage{ulem}
\usepackage{array}
\usepackage{tabularx}

\usepackage{amsmath}
\usepackage{amssymb}
\usepackage{xcolor}

\newcommand {\rootsNN}  {\ensuremath{\sqrt{s_{{\rm{NN}}}}}}

\usepackage{multirow}

\begin{document}

\title{Transport-model investigation of scaling of the number of constituent quarks and the hadronic-partonic transition in Au+Au collisions}

\bigskip

\author{Li-Ke Liu}
\affiliation{Key Laboratory of Quark {\&} Lepton Physics (MOE) and Institute of Particle Physics, 
Central China Normal University, Wuhan 430079, China}

\author{Shusu Shi}
\email[Corresponding author, ]{shiss@ccnu.edu.cn}
\affiliation{Key Laboratory of Quark {\&} Lepton Physics (MOE) and Institute of Particle Physics, 
Central China Normal University, Wuhan 430079, China}

\date{\today}
\begin{abstract}

We investigate the elliptic flow ($v_2$) in Au+Au collisions at $\sqrt{s_{\text{NN}}} = 3.0$ and 4.5 GeV using both hadronic and partonic transport models, including JAM, SMASH, AMPT-Hadronic Cascade, and AMPT-String Melting.
At 3.0 GeV, the JAM model reproduces the number-of-constituent-quark (NCQ) scaling violation observed by STAR, as well as the particle ordering ($K^0_S > p > \pi^+$).
Model calculations of the centrality dependence indicate that the scaling violation mainly originates from hadronic interactions rather than spectator effects, while the rapidity dependence further constrains the mechanism of the scaling breaking and the underlying longitudinal dynamics.
At 4.5 GeV, partonic interactions in the AMPT-String Melting mode significantly enhance NCQ scaling, and turning off final-state hadronic rescattering further clarifies the scaling pattern, highlighting the increasing role of partonic degrees of freedom.
The energy dependence of the $p_T$-integrated $v_2$ is also examined within these models.

\end{abstract}

\maketitle

\section{Introduction}\label{sec.I}

One of the primary goals of high-energy heavy-ion collision experiments is to create and study a new state of strongly interacting matter, the Quark-Gluon Plasma (QGP), 
where quarks and gluons are no longer confined inside hadrons but interact collectively through strong interactions~\cite{Collins:1974ky, Heinz:2000bk, PHENIX:2004vcz, BRAHMS:2004adc, STAR:2005gfr, PHENIX:2006dpn}. 
Demonstrating the transition from normal hadronic matter to this deconfined phase is a central topic in the study of hot and dense QCD matter~\cite{tHooft:1977nqb, Bzdak:2019pkr}.
Among various experimental observables, the elliptic flow ($v_2$) quantifies the azimuthal anisotropy of particle emission in heavy-ion collisions.
It is defined as the second-order harmonic coefficient in the Fourier expansion of the azimuthal distribution relative to the reaction plane:
\begin{equation}
\frac{dN}{d\phi}
\sim \frac{1}{2\pi}
\left( 1 + \sum_{n=1}^{\infty} 2 v_{n} \cos \left[ n(\phi - \psi_{\rm RP}) \right] \right),
\end{equation}
where $\frac{dN}{d\phi}$ is the azimuthal distribution of produced particles, 
and $\phi$ and $\psi_{\rm RP}$ denote the azimuthal angle of each particle and the  reaction plane, respectively.
It has proven to be a sensitive probe of the early-stage pressure gradients and the collective behavior of the produced medium~\cite{Poskanzer:1998yz, Voloshin:2008dg}.
A large $v_2$ signal indicates strong interactions among constituents and efficient translation of the initial geometric anisotropy into final-state momentum anisotropy.
Moreover, the approximate scaling of $v_2$ with the number of constituent quarks (NCQ scaling) provides evidence that partonic degrees of freedom dominate the medium’s dynamics before hadronization~\cite{STAR:2005gfr, PHENIX:2004vcz, STAR:2015gge, STAR:2017kkh, Braun-Munzinger:2007edi, Lu:2006qn}. 

At sufficiently high collision energies, 
systematic studies of collectivity across quark flavors --- from light to multi-strange hadrons ($K^0_S$, $\phi$, $\Lambda$, $\Xi$, $\Omega$) and even charm hadrons --- 
have provided powerful evidence for the formation of a strongly interacting QGP~\cite{STAR:2015gge, STAR:2017kkh, ALICE:2013olq, ALICE:2014wao}.
As the beam energy decreases, the temperature and energy density may drop below the threshold needed for deconfinement, 
causing the collective flow to be generated primarily through hadronic interactions, 
and the breakdown of NCQ scaling is therefore expected.
The RHIC Beam Energy Scan (BES) program covers the collision energy from {\rootsNN} = 3.0 to 62.4~GeV, 
corresponding to a baryon chemical potential range of 750 to 73 MeV~\cite{phase_tran:2010, Luo:2020pef, Chen:2024vb}. 
Data from BES-I show that NCQ scaling persists down to {\rootsNN} $\simeq$ 7.7 GeV~\cite{STAR:2013cow, STAR:2013ayu, STAR:2015rxv, STAR:2012och, Shi:2012ba}. 
At the lowest BES-I energy of 3.0~GeV, STAR observed that NCQ scaling no longer holds for light hadrons~\cite{STAR:2021yiu, Lan:2022rrc}. 

Recently, the new STAR measurements~\cite{STAR:2025owm} for {\rootsNN} = 3.2--4.5 GeV suggest that a partial reappearance of NCQ scaling might occur, 
hinting at a possible re-emergence of partonic degrees of freedom in this intermediate energy range.
The newly observed onset of NCQ scaling in this energy range poses an important test for transport approaches, 
which must describe the interplay between hadronic scatterings, possible partonic phases, and the collective flow development consistently across energies. 
The recent study has investigated the NCQ scaling of $v_{2}$ at 3.0--7.7 GeV using A Multi-Phase Transport Model (AMPT)~\cite{Zhu:2025kud}.

In this work, we employ several well-established transport models, including Jet AA Microscopic Transport Model (JAM)~\cite{Nara:2021fuu, Nara:2022kbb}, 
Simulating Many Accelerated Strongly interacting Hadrons (SMASH)~\cite{SMASH:2016zqf}, and AMPT~\cite{PhysRevC.72.064901, Yong:2023uct}. 
Specifically, the simulations are performed with JAM version~2.1~\cite{Nara:2022kbb}, 
SMASH version~2.0~\cite{dmytro_oliinychenko_2020_4336358}, and AMPT (2021 release) incorporates the hadronic cascade mode based on the 2018 framework, 
to calculate $v_2$ in this beam energy region. 
By comparing their predictions, we assess the robustness of the observed NCQ scaling,  
and test how well current transport approaches can capture the evolution of collectivity and the transition between hadronic and partonic dynamics.
Therefore, the paper is structured as follows: we first outline the key features of the transport model and its application at low collision energies, then present the detailed calculations and comparison with experimental measurements, and finally discuss the implications for our understanding of the QCD phase structure at high baryon density.

\section{Transport Models}\label{sec.II}
Since the evolution of relativistic heavy-ion collisions involves a complex many-body quantum system, 
a first-principle calculation is not yet feasible~\cite{Heinz:2013th}.
Transport models provide an effective theoretical framework to describe the entire dynamical evolution of heavy-ion reactions from the early to the late stages~\cite{Aichelin:1991xy, Bass:1998ca}.

Hadronic transport approaches have been developed for more than three decades. 
JAM is a hadronic cascade that includes resonance production, string excitations with their decays, 
and optional mean-field potentials, making it well suited for intermediate and low collision energies~\cite{Nara:2021fuu, Nara:2022kbb}. 
Simulating Many Accelerated Strongly-interacting Hadrons (SMASH) is a modern, modular hadronic baseline emphasizing a consistent treatment of hadronic cross sections and resonance dynamics, with optional baryonic mean-field effects~\cite{SMASH:2016zqf, Zhou:2024cte}.

On the multi-phase side, the AMPT framework integrates fluctuating initial conditions (HIJING), a partonic cascade (ZPC), quark-coalescence hadronization, and a hadronic afterburner (ART)~\cite{PhysRevC.72.064901}. In addition to the widely used string-melting setup (AMPT-SM), which converts excited strings into partons when partonic degrees of freedom are relevant, the recently developed hadron-cascade mode (AMPT-HC) enables pure hadronic cascade simulations with hadronic mean-field potentials~\cite{Yong:2023uct}.

Because of the rare yield of anti-baryon at low collision energies, the baryon asymmetry leads to a significant mean-field effect. 
therefore, it is crucial to include the baryonic mean field. 
The equations of motions have to be adjusted according to the modified one-particle Hamiltonian $H_{i}$:
\begin{equation}
H_i=\sqrt{\vec{p}_i^2+m_{\mathrm{eff}}^2}+U\left(\vec{r}_i\right)
  \label{Eq:Hamiltonian_U}
\end{equation}

where $\vec{p}_i$ represent the momentum vector, $m_{\mathrm{eff}}$ is the mass for stable hadrons and the effective mass for resonances (determined by their mass distribution), 
and $U\left(\vec{r}_i\right)$ is the distance dependent potential. Thus, the potential depends only on particle coordinates, but not on their momenta. 
The corresponding Hamiltonian basic equations are:  
\begin{equation}
\begin{aligned}
\frac{d \vec{r}_i}{d t} & =\frac{\partial H_i}{\partial \vec{p}_i}=\frac{\vec{p}_i}{\sqrt{\vec{p}_i^2+m_{\mathrm{eff}}^2}} \\
\frac{d \vec{p}_i}{d t} & =-\frac{\partial H_i}{\partial \vec{r}_i}=-\frac{\partial U}{\partial \vec{r}_i}
\end{aligned}
  \label{Eq:Hamiltonian_d}
\end{equation}
That means momentum conservative is valid only on average. 
The potential is calculated as a function of the local density
\begin{equation}
U=a\left(\rho / \rho_0\right)+b\left(\rho / \rho_0\right)^\tau \pm 2 S_{\mathrm{pot}} \frac{\rho_{I 3}}{\rho_0}
  \label{Eq:MF_potential}
\end{equation}
here $\rho$ is the Eckart rest frame baryon density and $\rho_{I3}$ is the Eckart baryon isospin density relative isospin projection $I_3/I$. 
$\rho_0=0.168\ \mathrm{fm}^{-3}$ is the nuclear ground state density. 
In the symmetry term (third term), a positive sign is used for neutrons, and the opposite sign is applied to protons. 
This treatment is restricted to baryons, with mesons being excluded.
Electromagnetic interactions, such as the Coulomb and Lorentz forces, 
are neglected as their contribution is usually much weaker compared to the hadronic mean-field effects.

Parameters for the Skyrme potential applied in different models are listed in table~\ref{tab1}~\cite{Isse:2005nk, SMASH:2016zqf, Yong:2023uct}.
For both JAM and SMASH, we employ a soft equation of state (EoS) without momentum dependence, 
corresponding to a nuclear incompressibility of $K = 210$~MeV, since the hard EoS leads to a much more negative $v_2$.
In contrast, for AMPT, the hard EoS is required to reproduce the observed negative $v_2$ signal.

\begin{table}[htbp]
\centering
\caption{Mean-field parameters for different transport models.}
\label{tab1}
\begin{tabular}{lccc}
\hline
Transport Models & JAM  & SMASH & AMPT-HC \\
\hline
$a$ (MeV) & -356  & -356 & -124 \\
$b$ (MeV) & 303   & 303  & 70.5 \\
$\tau$        & $7/6$  & $7/6$  &2 \\
$S_{\text{Pot}}$ (MeV)    & 18  & 18 & 18 \\
\hline
\end{tabular}
\end{table}

At low energy collisions, the reduced Lorentz contraction due to slower projectile velocities allows spectator matter to pass through over a longer time, 
making its shadowing effect much more pronounced.
Therefore, a detailed evaluation of the passing time and nuclear shadowing effect is essential~\cite{Herrmann:1999wu, Liu:1998yc, Liu:2024ugr}.
The passing time of the projectile and target spectators in the center of mass (CM) frame can be estimated as

\begin{equation}
t_{\text{pass}} \sim \frac{2R}{\gamma \beta},
\end{equation}

where $R$ = 6.98 fm is the radius of a gold nucleus, $\beta = v/c$ denotes the velocity of the nuclei in the CM frame normalized to the speed of light $c$, 
and $\gamma = 1/\sqrt{1-\beta^{2}}$ is the Lorentz factor corresponding to $\beta$.
The passing times of nuclear spectator for typical Au+Au collisions at RHIC energies are listed in~\ref{tab2}.
As the collision energy increases from {\rootsNN} = 3.0 to $4.5$~GeV, the spectator velocity rises from about $0.78$ to $0.91~c$, and the passing time decreases by roughly $40\%$. 
This faster separation shortens the shadowing period over the participant zone, 
weakening the out-of-plane expansion at lower energies and allowing in-plane elliptic flow to develop more rapidly.
In the transport simulations, such as SMASH, 
the spectator shadowing effect is naturally accounted for by propagating all nucleons in the self-consistent nuclear geometry. 
At each time step, the positions and momenta of individual nucleons are updated, so that particles emitted from the participant zone may be absorbed or deflected if their trajectories intersect with the space–time region occupied by spectators~\cite{SMASH:2016zqf}.

\begin{table}[htbp]
\centering
\caption{Passing time of nuclear spectator for typical Au+Au collisions at RHIC energies}
\label{tab2}
\begin{tabular}{lccccc}
\hline
Collision Energy (GeV) & 3.0 & 3.2 & 3.5 & 3.9 & 4.5 \\
\hline
Velocity & 0.78  & 0.81 & 0.85 & 0.88  & 0.91 \\
$t_{\rm pass}$ ($\text{fm}/c$) & 11.22  & 10.02  & 8.73 & 7.59 & 6.43 \\
\hline
\end{tabular}
\end{table}

\begin{figure*}[!htbp]
  \centering   
  \includegraphics[width=0.8\textwidth]{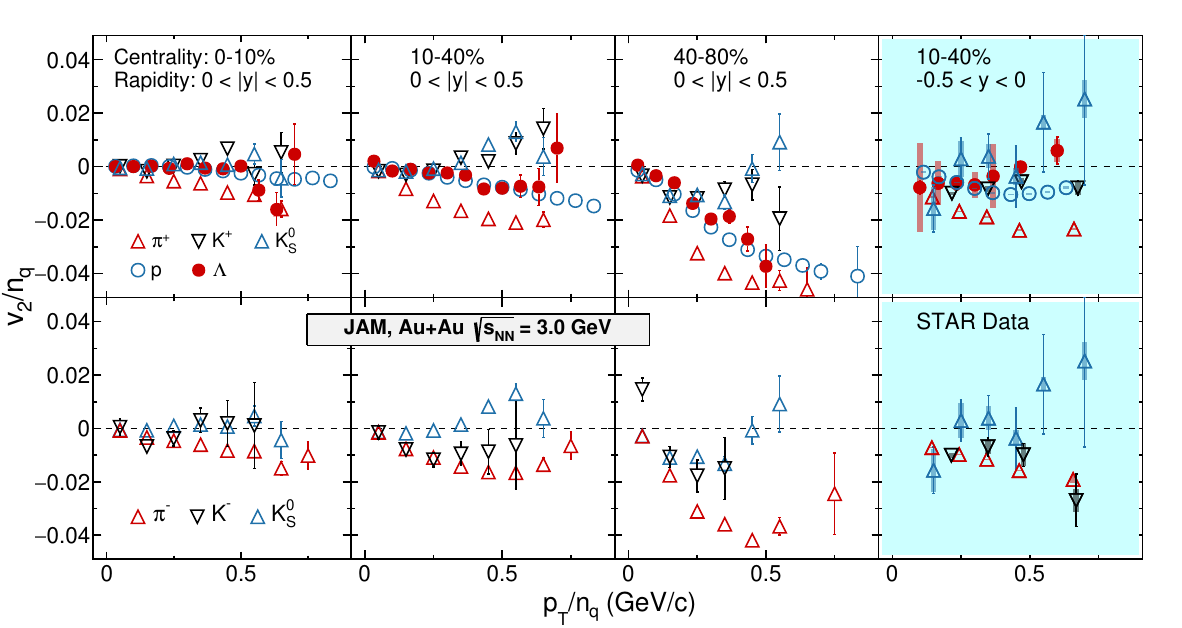}
  \includegraphics[width=0.8\textwidth]{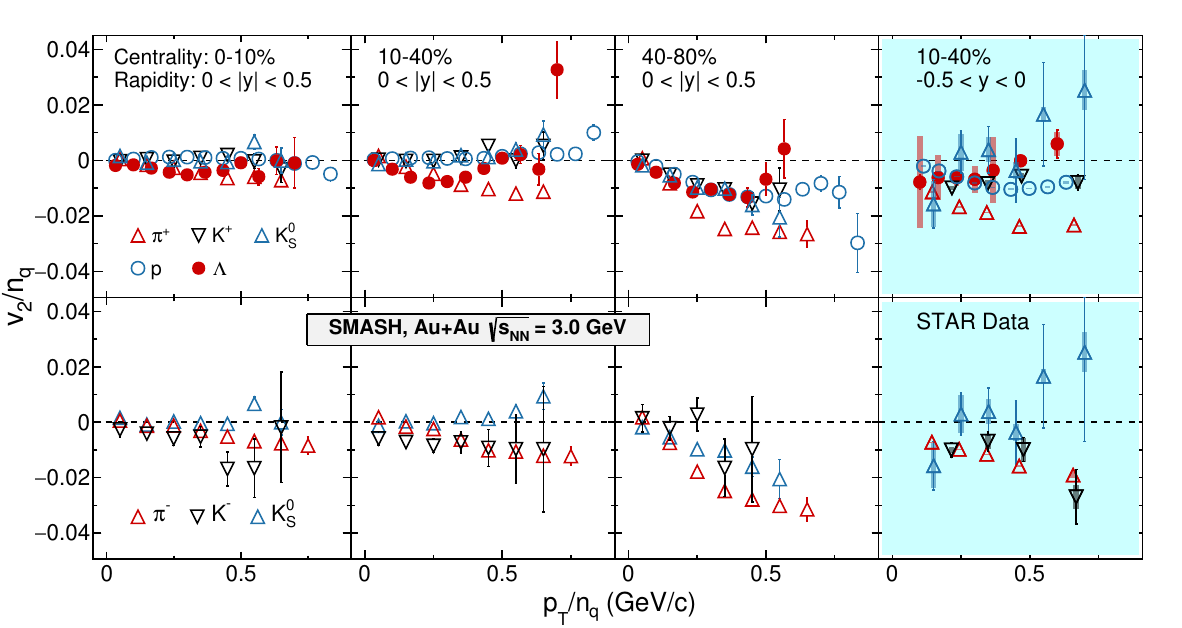}
  \includegraphics[width=0.8\textwidth]{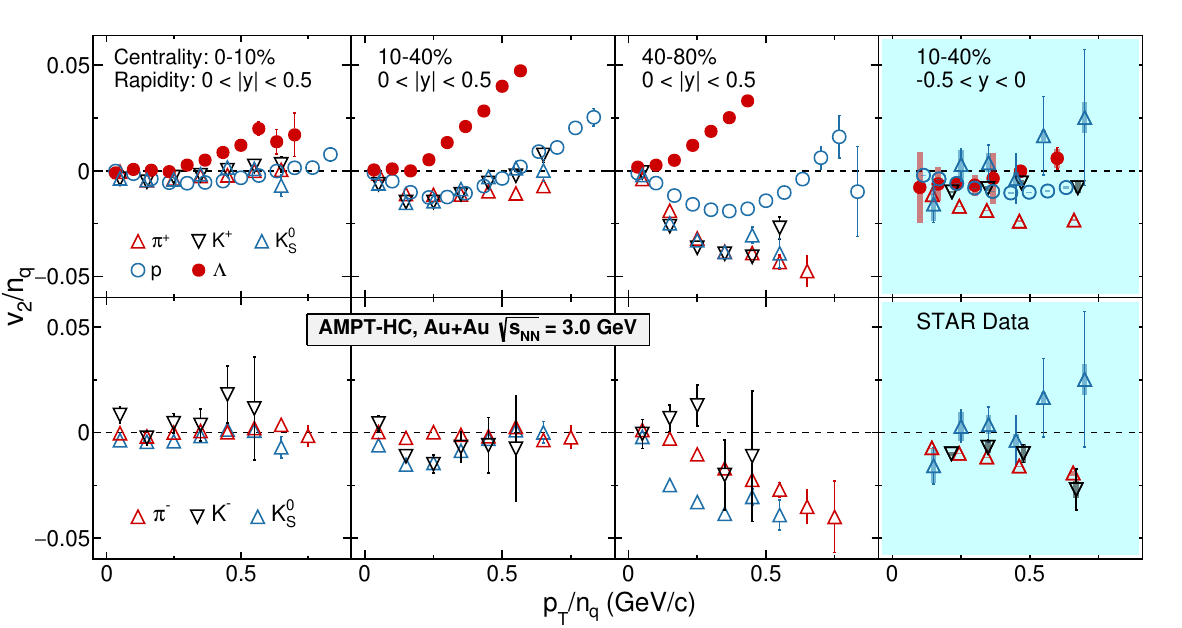}
   \caption{The number-of-constituent-quark ($n_q$) scaled elliptic flow ($v_{2}/n_{q}$) as a function of $n_q$-scaled transverse momentum ($p_{T}/n_q$) for 0--10\%, 10--40\%, and 40--80\% central Au+Au collisions at $\sqrt{s_{\text{NN}}} = 3.0$ GeV, obtained from model calculations with JAM, SMASH, and AMPT-HC, 
   compared with the STAR publication (right panel)~\cite{STAR:2021yiu, STAR:2025owm}.
   }
    \label{Fig1}
\end{figure*}

\begin{figure*}[htbp]
    \centering
    \includegraphics[width=0.66\textwidth]{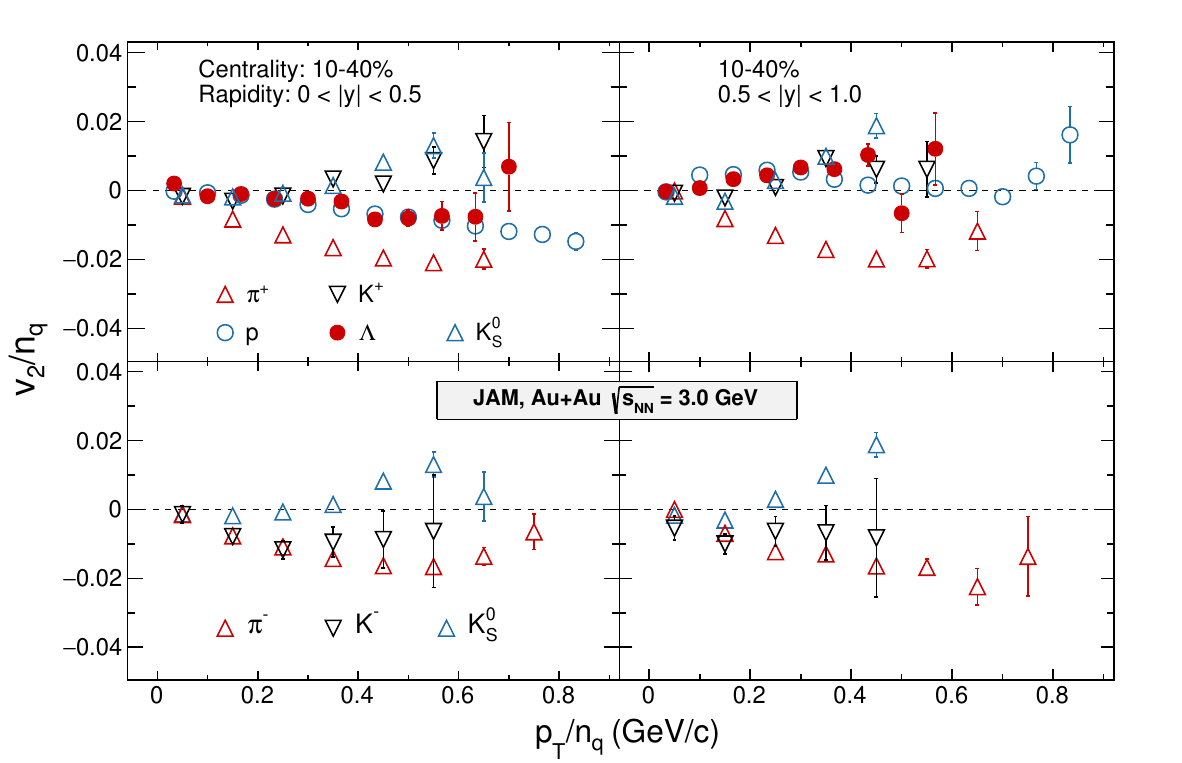}
    \includegraphics[width=0.66\textwidth]{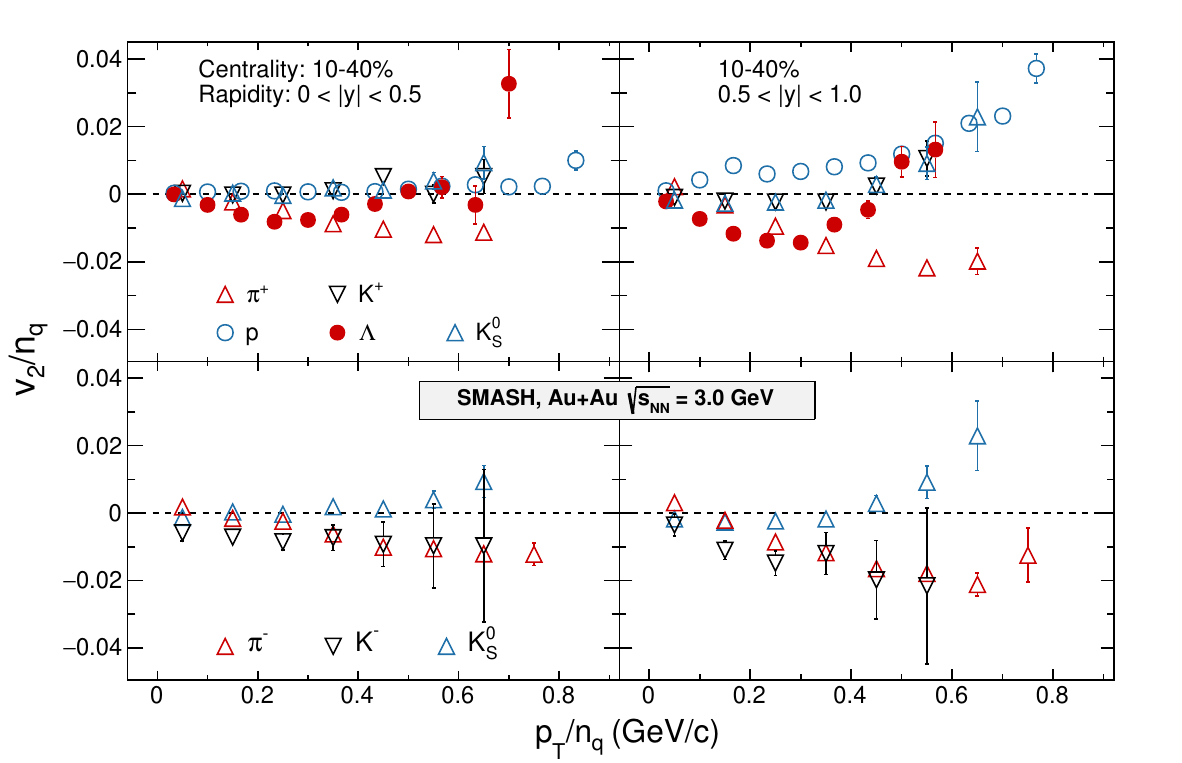}
    \includegraphics[width=0.66\textwidth]{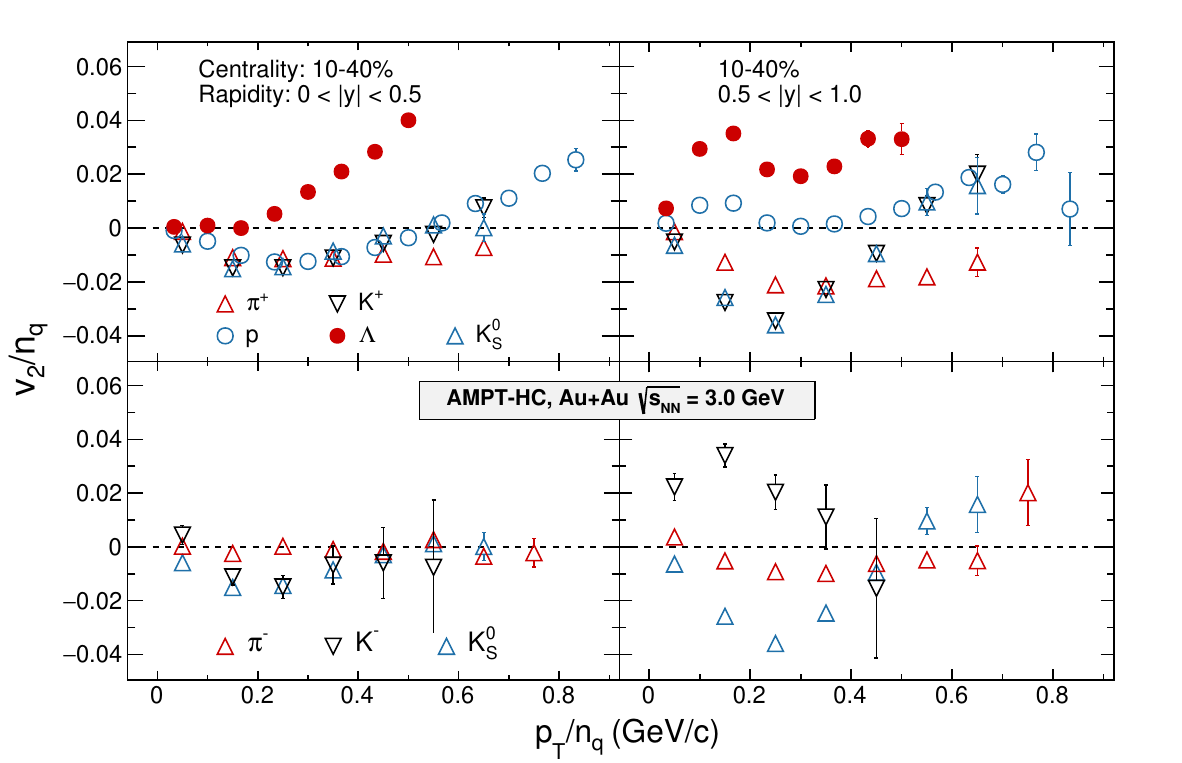}
    \caption{Number-of-constituent-quark ($n_q$) scaled elliptic flow ($v_{2}/n_{q}$) as a function of $n_q$-scaled transverse momentum ($p_{T}/n_q$) in Au+Au collisions at $\sqrt{s_{\text{NN}}} = 3.0$ GeV, obtained from JAM, SMASH, and AMPT model calculations for 10–40\% centrality at mid-rapidity ($0 < |y| < 0.5$) and forward rapidity ($0.5 < |y| < 1.0$).}
    \label{Fig2}
\end{figure*}

The new measurements on $v_{2}$ from STAR provide a clear picture of the energy-dependent role of spectator shadowing~\cite{STAR:2025owm}: 
at {\rootsNN} = 3.0~GeV, strong spectator blocking suppresses in-plane expansion, 
keeping $v_2$ negative for all charged hadrons, 
as the collision energy increase, shadowing effect reduce, and $v_{2}$ become more positive, closing 0 around 3.2 GeV.
The onset of partonic collectivity, along with reduced spectator shadowing, gives rise to positive $v_2$ and a better realization of NCQ scaling.
These conclusions are supported qualitatively by comparisons to several transport models~\cite{STAR:2025owm}.
Hadronic transport models well catch the negative $v_2$ signal and NCQ scaling breakdown,
and the multiphase model which include the partonic interactions reproduce the NCQ scaling.

In this work, we extend the investigation by performing a systematic study of NCQ scaling in Au+Au collisions 
over different centralities and rapidity ranges, to test the influence of initial geometry and spectators shadowing. 
We examine how hadronic interactions alone can lead to the observed scaling violation, 
and employ the AMPT string melting mode to verify that the inclusion of partonic degrees of freedom is the primary driver for restoring NCQ scaling. 
Finally, we present the energy dependence of the integrated $v_2$ to further quantify the transition from hadronic to partonic collectivity.

\section{Results and Discussions}\label{sec.III}

\begin{figure*}[htbp]
    \centering
    \includegraphics[width=0.95\textwidth]{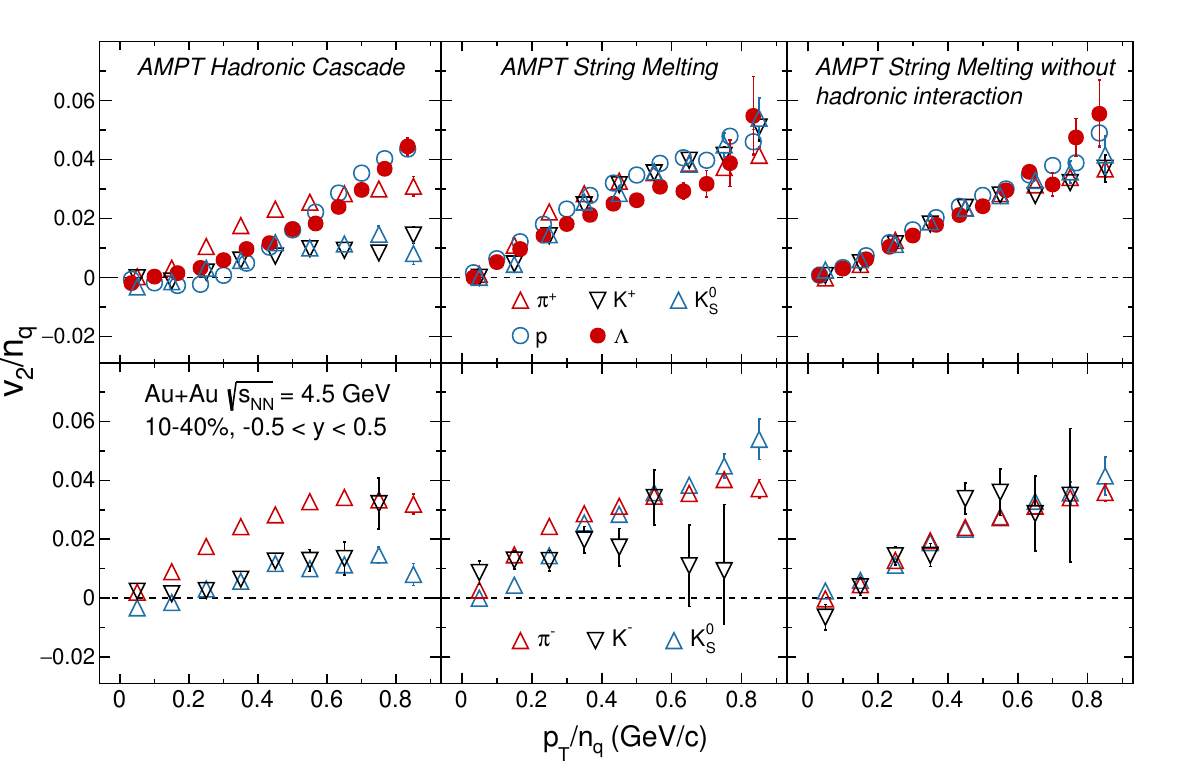}
    \caption{Number-of-constituent-quark ($n_q$) scaled elliptic flow ($v_{2}/n_{q}$) as a function of $n_q$-scaled transverse momentum ($p_{T}/n_q$) for Au+Au collisions at $\sqrt{s_{\text{NN}}} = 4.5$ GeV, obtained from AMPT model calculations in three configurations: Hadronic Cascade, String Melting, and String Melting without hadronic interactions.}
    \label{Fig3}
\end{figure*}

\begin{figure*}[!htb]
    \centering
    \includegraphics[width=0.8\textwidth]{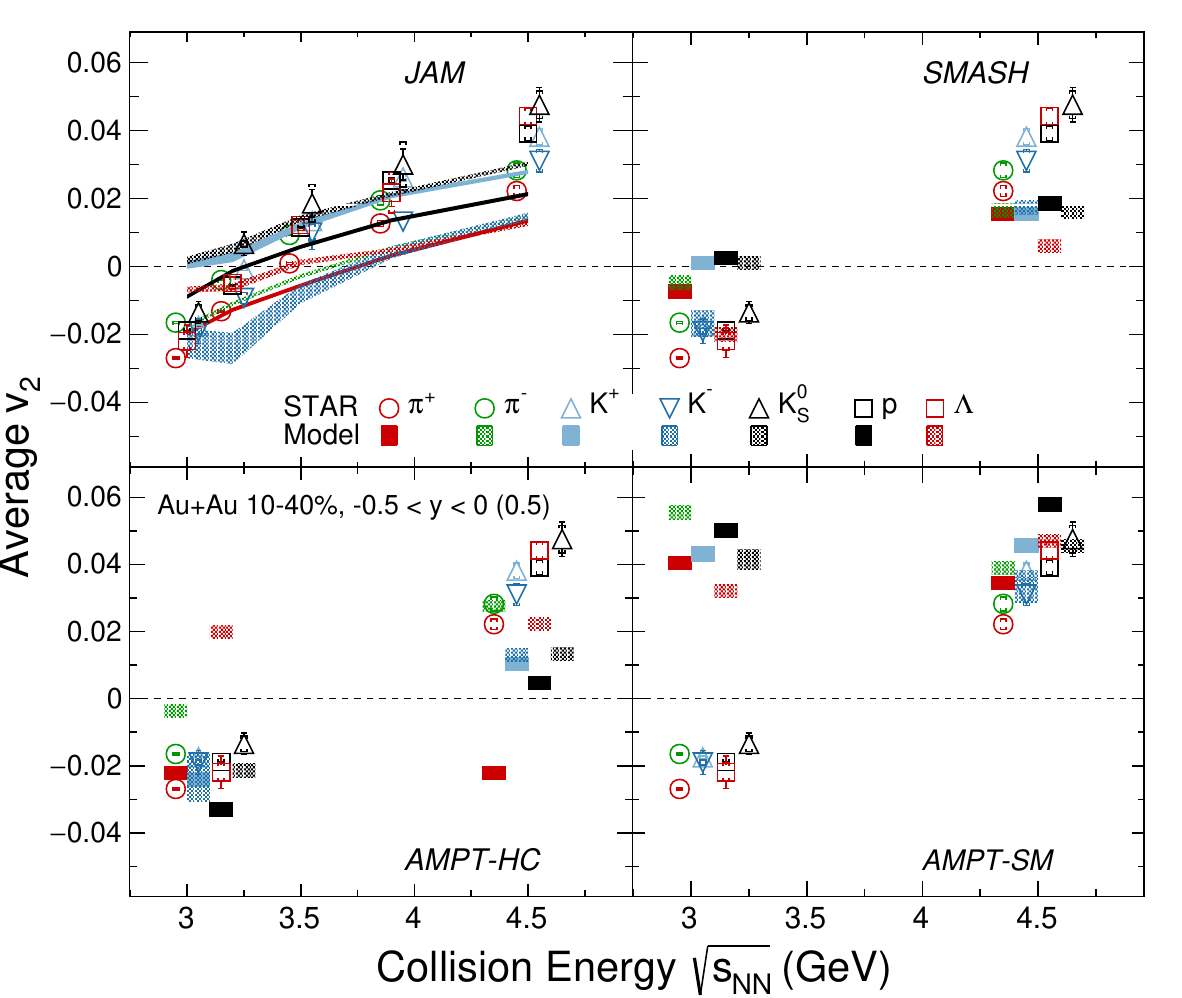}
    \caption{Integrated elliptic flow ($v_2$) of $\pi^{+}$, $K^{+}$, $K^{0}_{S}$, $p$, and $\Lambda$ in 10--40\% central Au+Au collisions at $\sqrt{s_{\text{NN}}} $ = 3.0--4.5~GeV, extracted from various transport model calculations. Model results from JAM, AMPT-HC, AMPT-SM, and SMASH are shown as bands. Experimental data points are taken from STAR publications~\cite{STAR:2021yiu, STAR:2025owm}.
    For clarity, the x-axis values of integrated $v_2$ are shifted a little.}
    \label{Fig4}
\end{figure*}

All model calculations are performed in the CM frame.
The collision centrality is determined by classifying events according to the percentile of particle multiplicity, using only charged hadrons ($\pi^{\pm}$, $K^{\pm}$, $p$, and $\bar{p}$), consistent with the experimental procedure.
Since the hadronic transport models are well suited for describing heavy-ion collisions at $\sqrt{s_{\text{NN}}} = 3.0$ GeV, 
thus, we employ JAM, SMASH, and AMPT-HC to investigate the centrality dependence of $v_{2}$ and the impact of spectator shadowing at forward rapidity. 
Figure~\ref{Fig1} shows the number-of-constituent-quark ($n_q$) scaled $v_{2}$ as a function of $n_q$-scaled transverse momentum in 0--10\%, 10--40\%, and 40--80\% central Au+Au collisions at $\sqrt{s_{\text{NN}}} = 3.0$ GeV, based on model calculations from JAM, SMASH, and AMPT-HC.
Notably, the JAM calculation for 10--40\% centrality not only reproduces the violation of NCQ scaling but also captures the observed particle ordering of $v_{2}$ ($K^{0}_{S} > p > \pi^{+}$), 
demonstrating that the differential collective flow is primarily governed by hadronic cross sections.

In more peripheral collisions, the larger initial geometric eccentricity would normally enhance $v_{2}$; 
however, spectator effects significantly suppress the in-plane expansion, leading to more negative $v_{2}$ signals. 
Since the centrality dependence of $v_{2}$ essentially reflects the role of spectator effects, studying different centralities provides a direct way to assess their impact. 
Our multi-model calculations show that NCQ scaling is violated at all centralities, including 0--10\%, consistent with the expectation that hadronic transport models, which do not include partonic collectivity, generally do not exhibit NCQ scaling. 
This suggests that the observed scaling violation primarily originates from hadronic interactions, with larger initial eccentricity and spectator shadowing further influencing the overall behavior.

At forward rapidity, the spectator shadowing effect is expected to be reduced, causing $v_{2}$ to shift toward positive values.
In addition, the density of the interaction zone decreases, 
so the effective partonic and hadronic interactions are also weakened. 
Measurements of $v_{2}$ at forward rapidity therefore provide sensitivity 
to the role of nuclear shadowing. 

Figure~\ref{Fig2} shows the $n_q$-scaled elliptic flow ($v_{2}/n_{q}$) as a function of $n_q$-scaled transverse momentum ($p_{T}/n_q$) for Au+Au collisions at $\sqrt{s_{\text{NN}}} = 3.0$ GeV, calculated with JAM, SMASH, and AMPT, for 10–40\% centrality at mid-rapidity ($0 < |y| < 0.5$) and forward rapidity ($0.5 < |y| < 1.0$).
Comparing mid-rapidity and forward rapidity, the $v_{2}$ values of $p$ and $\Lambda$ become more positive at forward rapidity in JAM. 
This behavior can be attributed to weaker spectator effects at forward rapidity: since spectator matter tends to block in-plane emission, 
its reduction allows $v_{2}$ to shift toward positive values. 
Moreover, at low beam energies, the production of $\Lambda$ hyperons is dominated by associated production processes (e.g., $NN \to N\Lambda K$), which naturally link their collective behavior to that of protons, leading to similar flow patterns.
Pions, however, show little variation with rapidity, likely because their large hadronic cross section maintains strong rescattering even at forward rapidity.
The results from SMASH and AMPT-HC both show a stronger violation of NCQ scaling.
Therefore, future $v_2$ measurements at forward rapidities at facilities such as FAIR-CBM, JINR-NICA, and HIRFL/HIAF-CEE are highly warranted~\cite{Lu:2016htm, Liu:2023xhc, Zhang:2023hht, Kisiel:2020spj, MPD:2025jzd, CBM:2016kpk, CBM:2025voh}.
Such measurements can help disentangle the different mechanisms responsible for the breaking of NCQ scaling and shed light on the underlying longitudinal dynamics.

We investigate the roles of hadronic and partonic interactions in Au+Au collisions at $\sqrt{s_{\text{NN}}} = 4.5$ GeV by applying different modes of the AMPT model: Hadronic Cascade (HC), String Melting (SM), and String Melting without final-state hadronic interactions, as shown in Fig.~\ref{Fig3}.
In the pure HC mode, no NCQ scaling is observed, as all scatterings occur only between hadrons.
With partonic interactions included in the SM mode, the scaling behavior is significantly improved, indicating that partonic scatterings and quark coalescence contribute substantially to the observed scaling.
Moreover, when final-state hadronic rescattering is switched off and only partonic interactions are retained, the NCQ scaling becomes even more pronounced, as shown in the right panel of Fig.~\ref{Fig3}, indicating that a strong scaling behavior emerges when partonic degrees of freedom dominate the system’s evolution.
These results demonstrate that NCQ scaling is a sensitive probe of partonic dynamics, and are consistent with the development of partonic collectivity in heavy-ion collisions.

Figure~\ref{Fig4} presents the energy dependence of $p_T$-integrated $v_2$ in 10–40\% central Au+Au collisions at $\sqrt{s_{\text{NN}}} = 3.0$ and 4.5 GeV. The data are taken from STAR publications~\cite{STAR:2021yiu, STAR:2025owm}. 
Model results are calculated within the same $p_T$ range with the $p_T$ spectra weighting and
a symmetric rapidity window ($-0.5<y<0.5$) for better statistics, exploiting the symmetry of $v_2$ about midrapidity.

JAM reproduces the 3.0 GeV results reasonably well and captures the general trend of the energy dependence, likely because it accounts for the spectator shadowing effect, which decreases over the passing time of the nuclei. 
However, JAM significantly underestimates $v_2$ at 4.5 GeV. Similarly, SMASH and AMPT-HC partially reproduce the negative $v_2$ at 3.0 GeV, but also underestimate the 4.5 GeV results.
The AMPT-SM mode provides a better description of the $v_2$ data at 4.5 GeV, suggesting that partonic interactions are essential for reproducing the observed $v_2$ at this energy.
All integrated $v_2$ values at 3.0 GeV from AMPT-SM remain positive, likely resulting from the omission of the finite nuclear thickness in the model, which diminishes the spectator shadowing effect~\cite{Lin:2025private}.

\section{Summary}\label{sec.IV}
Our study investigates elliptic flow ($v_2$) in Au+Au collisions over a range of energies and centralities using hadronic and partonic transport models. 
At $\sqrt{s_{\text{NN}}} = 3.0$ GeV, hadronic transport models (JAM, SMASH, AMPT-HC) reproduce many features of the data.  
JAM, in particular, captures the violation of NCQ scaling and the observed ordering of $v_2$ ($K^{0}_{S} > p > \pi^{+}$), 
indicating that differential collective flow is largely determined by hadronic cross sections. 
The centrality dependence reflects the influence of spectator shadowing, 
with forward rapidity showing weaker spectator effects and correspondingly more positive $v_2$ for heavier hadrons, 
while pions are less sensitive due to their large cross sections. 
Across all centralities, NCQ scaling is violated, consistent with the expectation for hadronic transport models that do not include partonic collectivity.

At higher energy, $\sqrt{s_{\text{NN}}} = 4.5$ GeV, AMPT calculations reveal the roles of partonic interactions. 
In the pure Hadronic Cascade mode, NCQ scaling is absent, whereas the inclusion of partonic scatterings and quark coalescence in the String Melting (SM) mode significantly improves the scaling. 
Turning off final-state hadronic rescattering further enhances the scaling, demonstrating that partonic degrees of freedom play a dominant role in establishing NCQ scaling. 
These results support the use of NCQ scaling as a sensitive probe of partonic collectivity in heavy-ion collisions.
The energy dependence of $p_T$-integrated $v_2$ further highlights the interplay between hadronic and partonic effects. 
JAM reproduces the 3.0 GeV $v_2$ and captures the general trend with energy, likely due to its treatment of spectator shadowing, but underestimates $v_2$ at 4.5 GeV. 
SMASH and AMPT-HC partially reproduce the negative $v_2$ at 3.0 GeV but also underestimate the higher-energy results. 
In contrast, the AMPT-SM mode better describes the 4.5 GeV $v_2$, whereas all integrated $v_2$ values at 3.0 GeV remain positive, 
possibly due to the omission of finite nuclear thickness and reduced spectator shadowing in the model.

While at higher energies, partonic scatterings and quark coalescence become increasingly important in shaping both the NCQ scaling and the magnitude of elliptic flow, measurements of centrality and rapidity dependence are crucial for disentangling the breaking of NCQ scaling from spectator effects.
In the future, high-statistics data from FAIR-CBM, JINR-NICA, and HIRFL/HIAF-CEE in the high-baryon-density regime will provide valuable opportunities. The flow of multi-strange hadrons and $\phi$ mesons can offer new insights into partonic collectivity, while studies of atomic-number scaling of elliptic flow will shed light on the production mechanisms of light and hypernuclei~\cite{Yan:2006bx, Xu:2025unm, Zhao:2025glf, Zhou:2025zgn}.
In addition, studies of smaller systems at higher energies are essential to clarify the energy and system-size dependence of deconfinement~\cite{Huang:2019tgz, Ambrus:2024eqa}.
On the modeling side, improvements to AMPT-SM are needed, particularly for heavy-ion collisions in the high-baryon-density region, including the implementation of hadronic mean-field effects and the treatment of finite nuclear thickness.

{\bf Acknowledgments:}
We are grateful for discussions with Prof. Zi-Wei Lin.
This work is supported in part by the National Key Research and Development Program of China under Contract Nos. 2022YFA1604900 and 2024YFA1610700; the National Natural Science Foundation of China (NSFC) under contract No. 12175084.

{\bf{Data availability:}}
The data that support the findings of this article are not publicly available. 
The data are available from the authors upon reasonable request.

\bibliography{example} 

\end{document}